# PRIME: Phase Reversed Interleaved Multi-Echo acquisition enables highly accelerated distortion-free diffusion MRI

Yohan Jun[1,2,3], Qiang Liu[4,5], Ting Gong[1,2], Jaejin Cho[6], Shohei Fujita[1,2], Xingwang Yong[1,2,7], Congyu Liao[8], Marianna E Schmidt[9,10], Shahin Nasr[1,2], Camilo Jaimes[2,3,12], Michael S Gee[2,3,11], Susie Y Huang[1,2,12], Lipeng Ning[3,13], Anastasia Yendiki[1,2], Yogesh Rathi[2,13,14], and Berkin Bilgic[1,2,12]

[1]Athinoula A. Martinos Center for Biomedical Imaging, Massachusetts General Hospital, Boston, MA, United States
[2]Department of Radiology, Harvard Medical School, Boston, MA, United States
[3]Pediatric Imaging Research Center, Massachusetts General Hospital, Boston, MA, United States
[4]Brigham and Women's Hospital, Harvard Medical School, Boston, MA, United States
[5]Department of Electrical and Computer Engineering, Northeastern University, Boston, MA, United States
[6]Department of Artificial Intelligence and Robotics, Sejong University, Seoul, Korea
[7]Zhejiang University, Hangzhou, China
[8]Department of Radiology and Biomedical Imaging, University of California, San Francisco, CA, United States
[9]Max Planck Institute for Human Cognitive and Brain Sciences, Leipzig, Germany
[10]Max Planck School of Cognition, Leipzig, Germany
[11]Department of Radiology, Massachusetts General Hospital, Boston, MA, United States
[12]Harvard/MIT Health Sciences and Technology, Cambridge, MA, United States
[13]Department of Psychiatry, Brigham and Women's Hospital, Boston, MA, United States
[14]Department of Radiology, Brigham and Women's Hospital, Boston, MA, United States

The word count of the body of the text: ~5000

Corresponding Author

Yohan Jun

Athinoula A. Martinos Center for Biomedical Imaging, Building 149, 13th Street, Rm 2301, Charlestown, MA 02129, USA.

Email: yjun@mgh.harvard.edu

Grant Support: This work was supported by research grants NIH R01 EB032378, R01 EB028797, R01MH132610, R01MH125860, R01MH116173, R01HD114719, R01EY030434, P41 EB030006, U01 EB026996, UG3 EB034875, R21 AG082377, R21 EB036105, NVIDIA Corporation for computing support, and National Research Foundation of Korea (NRF) grant funded by the Korea government (MSIT) (RS-2025-00555277). MES received funding from Max Planck School of Cognition, supported by the German Federal Ministry of Education and Research (BMBF) and the Max Planck Society.

Running Title: PRIME enables highly accelerated distortion-free diffusion MRI

**Abstract**

**Purpose:** To develop and evaluate a new pulse sequence for highly accelerated distortion-free diffusion MRI (dMRI) by inserting additional echoes without prolonging TR, when generalized slice dithered enhanced resolution (gSlider) radiofrequency encoding is used for volumetric acquisition.

**Methods:** A phase-reversed interleaved multi-echo acquisition (PRIME) was developed for rapid, high-resolution, and distortion-free dMRI, which includes several echoes where the first echo is for target diffusion-weighted imaging (DWI) acquisition with high-resolution and additional echoes are acquired with either lower resolution for 1) high-fidelity field map estimation, 2) phase navigation for shot-to-shot phase correction, 3) motion navigation across diffusion directions, or with high resolution to enable 4) high fidelity diffusion relaxometry acquisitions. The sequence was evaluated on *in vivo* data acquired from healthy volunteers on clinical and Connectome 2.0 scanners.

**Results:** *In vivo* experiments demonstrated that 1) high in-plane acceleration ($R_{in-plane}$ of 5-fold with 2D partial Fourier) was achieved using the high-fidelity field maps estimated from the second echo, which was made at a lower resolution/acceleration to increase its SNR while matching the effective echo spacing of the first readout, 2) high-resolution diffusion relaxometry parameters were estimated from triple-echo PRIME data using a white matter model of multi-TE spherical mean technique (MTE-SMT), and 3) high-fidelity mesoscale DWI at 490 μm isotropic resolution was obtained *in vivo* by capitalizing on the high-performance gradients of the Connectome 2.0 scanner.

**Conclusion:** The proposed PRIME sequence enabled highly accelerated, high-resolution, and distortion-free dMRI using additional echoes without prolonging scan time when gSlider encoding is utilized.

**INTRODUCTION**

Diffusion MRI (dMRI) has been widely used in neuroscience studies for analyzing *in vivo* brain tissue microstructure and neurological or neurodegenerative disorders, including stroke, white matter (WM) disorders, Alzheimer's disease, demyelination, tumors, and Parkinson's disease (1–7). However, in clinical settings, dMRI is typically acquired with relatively low resolution, e.g., about 2 mm isotropic resolution, compared to structural contrast-weighted images (8,9), which limits analyzing the detailed brain structures such as gray-WM boundary and white-matter fascicles in small cortical regions (8,10,11). Moreover, microstructure features of the human brain, including cell bodies or neurites, can be extracted from dMRI using diffusion models (4,12); however, the dMRI resolutions for the models are mostly limited to 1.5~2.5 mm isotropic resolution for *in vivo* analysis (13–18), which inevitably requires indirect estimation of microanatomical information beyond the voxel resolution and may depend on model assumptions and parameter settings (12). Nonetheless, increasing the resolution of dMRI is still challenging due to 1) low signal-to-noise ratio (SNR) coming from decreased voxel size and $T_2$ signal decay with long echo time (TE), 2) $T_2$ and $T_2^*$ voxel blurring because of lengthy echo spacing (ESP) and k-space readouts of echo-planar imaging (EPI), and 3) $B_0$ inhomogeneity and eddy current-induced geometric distortion (19–21).

Undersampling the k-space of dMRI with parallel imaging can mitigate signal decay, voxel blurring, and geometric distortion (22,23); however, it leads to residual aliasing artifacts and noise amplification with high in-plane acceleration factors. To overcome the limitations of the single-shot EPI for dMRI acquisitions, multi-shot EPI has become a popular approach (24–28), which enables the reduction of the $T_2$ and $T_2^*$ voxel blurring and geometric distortion by reducing the readout duration in each shot. However, combining multiple shots is difficult due to shot-to-shot phase variations coming from physiological noise. Several navigator-based (26,27,29–31) and navigator-free approaches (21,24,32–36), including multiplexed sensitivity encoding (MUSE) (24), multi-shot sensitivity-encoded diffusion data recovery using structured low-rank matrix completion (MUSSELS) (34), and low-rank modeling of local k-space neighborhoods (LORAKS) (37), have been proposed to combine multiple shots considering phase variations between the shots. In particular, the MUSE technique enabled a robust multi-shot EPI for high-resolution diffusion-weighted imaging (DWI), up to $0.3 \times 0.3 \times 8$ mm$^3$ resolution for b = 500 s/mm$^2$ (24), and has been used for clinical practice and validation (38–42).

Moreover, to overcome the SNR limitation of 2D EPI-based acquisitions, 3D EPI techniques for volumetric acquisition, such as 3D multi-slab acquisitions (43–48), have been proposed. Engstrom et al (43) proposed 3D multi-slab acquisition with 2D phase navigation, enabling up to 1.3 mm isotropic resolution for b = 1000 s/mm$^2$ with high fidelity. Generalized SLIce Dithered Enhanced Resolution Simultaneous Multi-Slab (gSlider-SMS) (10) is another acquisition technique that uses a slab radiofrequency (RF) encoding to acquire multiple volumetric slabs simultaneously.

Geometric distortions of dMRI due to B$_0$ inhomogeneity and eddy currents can be corrected by obtaining additional shots with reversed phase-encoding and estimating field maps using software, such as FSL TOPUP (49,50). Hybrid-space SENSitivity Encoding (SENSE) has been proposed for joint reconstruction of interleaved blip-up and -down images with incorporated field maps estimated from FSL TOPUP, which showed improved g-factor compared to standard parallel imaging reconstruction for each of the blip-up and -down shots separately (51,52). Recently, another approach, Echo-Planar Time-Resolved Imaging (EPTI) (53), was developed for distortion-free dMRI along with relaxometry imaging using spatiotemporal encoding and reconstruction schemes (54,55), which was further extended to mesoscale resolution using ROMER acquisition (56). BUDA-EPI (21), which is a blip-up and -down acquisition (BUDA) (57) and joint parallel imaging reconstruction with a structured low-rank constraint framework (37), enabled high-resolution dMRI with high-fidelity by combining it with a gSlider acquisition technique, and BUDA-circular-EPI (20) further pushed the resolution by reducing the readout length and TE with circular EPI acquisition. However, there are limitations of BUDA-EPI. First, achievable acceleration per shot is limited since residual artifacts in the interim SENSE reconstructed images can propagate to the estimated field maps using FSL TOPUP, which constrains BUDA-EPI to use R$_{in-plane}$ of 4-fold or lower. Second, the long repetition time (TR) needed for gSlider RF encoding to mitigate spin history-related slab boundary artifacts and slab-cross talk effects (10) leaves significant dead time, especially when simultaneous multi-slice (SMS) (58) is employed.

To address these drawbacks, we propose a new pulse sequence for highly accelerated distortion-free dMRI, called PRIME (**P**hase **R**eversed **I**nterleaved **M**ulti-**E**cho acquisition). PRIME benefits from inserting additional echoes without incurring a scan time penalty when gSlider RF encoding is used. The second or third echo can be utilized in two different ways: 1) high-fidelity field maps estimation along with shot-to-shot phase and motion navigator, and 2) diffusion relaxometry acquisition. We demonstrated that highly accelerated, high-resolution, and distortion-free dMRI was achieved using the PRIME sequence on clinical and Connectome 2.0 scanners (59,60). While there were studies that utilized second echo acquisition as a phase navigator for shot-to-shot phase correction with (NAViEPI) (61) or without (62) matching effective ESP between echoes, estimating high-fidelity field maps from additional echo acquisition to reconstruct high-fidelity target DWI and additionally utilizing the additional echo for diffusion modeling and motion estimation remains unexplored. The main contributions of our work are as follows:

- High in-plane acceleration, $R_{in\text{-}plane}$ of 5-fold with 2D partial Fourier (pF) for both phase encoding and frequency encoding directions, was achieved using the high-fidelity field maps estimated from the second echo, which is made at a lower resolution and acceleration to increase its SNR while matching the effective ESP of the first readout. This ensures that both readouts have identical geometric distortion, but also allows the second readout to use lower $R_{in\text{-}plane}$ so that the field maps are estimated from higher-quality interim reconstructions. High acceleration achieved in the first echo helped reduce relaxation-related voxel blurring with reduced ESP and TE, while joint reconstruction of blip-up and -down shots using low-rank constraint with phase prior (S-LORAKS) (37,63) showed improved reconstruction performance.
- Acquisition of the second echo enabled 1) rapid SENSE-based multi-shot DWIs reconstruction with shot-to-shot phase correction as a phase navigator, which is useful for reconstructing high-resolution mesoscale dMRI with a large number of diffusion directions, and 2) motion correction between diffusion directions as a motion navigator.
- High-resolution diffusion relaxometry parameters were estimated from triple-echo PRIME data, in which the second and third echoes have the same resolution, ESP, and $R_{in\text{-}plane}$ as the first echo to match the voxel size and distortion between echoes. This also benefits from long TR when gSlider RF encoding is used with SMS, where additional echoes can be acquired for diffusion relaxometry estimation without prolonging scan time.
- High-fidelity mesoscale *in vivo* dMRI of 490 μm isotropic resolution for b = 1,000 s/mm$^2$ was demonstrated by capitalizing on the high gradient performance of the Connectome 2.0 scanner (59,60), which has a maximum gradient strength of 500 mT/m and maximum slew rate of 600 T/m/s.

- The PRIME sequence was designed and implemented using the open-source MR sequence development framework (*Pulseq*) (64). The source codes for the sequence and reconstruction can be found here: https://github.com/yohan-jun/PRIME

## METHODS

### Pulse Sequence

Fig. 1a illustrates the pulse sequence diagram of PRIME. It features two or three spin echoes where the first echo is for target DWI acquisition with high resolution, and the second or third echo can be acquired with either 1) low-resolution for field map estimation along with phase and motion navigator, or 2) high-resolution for the second or third echo of target DWI acquisition. The effective ESP (defined as nominal ESP divided by $R_{in-plane}$) of the first and second or third echo readouts is matched to have the same geometric distortion. gSlider RF encoding is used to acquire multiple slabs and perform super-resolution reconstruction to resolve thin slices for high isotropic resolution dMRI (10,21), which is combined with blipped-controlled aliasing in parallel imaging (blipped-CAIPI) SMS technique (10,58). Blip-up and -down shots with reversed phase-encoding directions are acquired for each diffusion direction to correct geometric distortions of dMRI coming from $B_0$ inhomogeneity and eddy currents. Based on the gSlider RF encoding Toolbox (https://github.com/wgrissom/gSliderRF) (65), five gSlider RF encoding pulses are deployed, where 10 shots (5 gSlider RF encodings × blip-up/-down shots) are acquired per volume. Seven gSlider RF encoding pulses are deployed for mesoscale dMRI acquisition, resulting in 14 shots per volume. Variable-rate selective excitation (VERSE) (66,67) technique is used to mitigate specific absorption rate (SAR) constraints. The open-source MR sequence development framework, *Pulseq* (https://github.com/pulseq/pulseq), is used for the sequence implementation (64).

Fig. 1b shows the sequence timing comparison between the conventional BUDA-EPI (21) and PRIME. With $R_{in-plane}$ of 5-fold at $1 \times 1$ mm$^2$, ESP = 0.70 ms, multi-band (MB) factor of 2, and 2D pF of 6/8 × 6/8 on a clinical scanner, single-echo acquisition of BUDA-EPI with gSlider RF encoding has long dead time in each TR, e.g., around 182 ms out of $TR_{slice}$ = 270 ms in this case, to mitigate spin history-related slab boundary artifacts. PRIME utilizes this time by inserting additional refocusing pulse and readouts without prolonging TR when gSlider RF encoding is used.

To demonstrate the signal level of the second echo of PRIME, a comparison between the second echo images of PRIME and the single-echo images separately acquired with the same TE was conducted, as shown in Supporting Information Figure S1. This shows the reconstructed DWIs for b = 0 and 1,000 s/mm$^2$ with $R_{\text{in-plane}} \times$ MB = 4 × 2 and 2D pF (6/8 × 6/8) at 1 × 1 × 5 mm$^3$ slab, where TE = 118 ms. The signal norm of the second echo image compared to the single-echo image was 0.97 and 1.01 for b = 0 and 1,000 s/mm$^2$, respectively, which demonstrates that signal levels of PRIME were comparable with those of single-echo images that were separately acquired with another scan.

**k-space Trajectory**

The representative schematic k-space trajectories of PRIME are presented in Fig. 2. For high-fidelity field maps estimation (Fig. 2a), the first echo is acquired with $R_{\text{in-plane}}$ of 5-fold at 1 × 1 mm$^2$ and 2D pF on a clinical scanner, whereas the second echo is acquired with the same effective ESP as the first echo for $R_{\text{in-plane}}$ of 4-fold at 2 × 2 mm$^2$ and 1D pF. High in-plane acceleration with 2D pF can mitigate relaxation-related voxel blurring with reduced ESP and TE. The blip-up and -down shots cover complementary k-space positions, and all quadrants of k-space are covered with the aid of the virtual coil constraints in S-LORAKS (37,63), which will be covered in the next section of *Image Reconstruction*.

For high-resolution diffusion relaxometry (Fig. 2b), all echoes have the same resolution of 1 × 1 mm$^2$ and $R_{\text{in-plane}}$ of 3-fold with 2D pF to match the voxel size and distortion. With the high gradient strength of the Connectome 2.0 scanner (59,60), mesoscale diffusion imaging (Fig. 2a) of 490 × 490 μm$^2$ resolution $R_{\text{in-plane}}$ of 4-fold with 2D pF can be achieved, where the second echo has the resolution of 2 × 2 mm$^2$ and $R_{\text{in-plane}}$ of 2-fold with 2D pF.

**Image Reconstruction**

Fig. 3a demonstrates the standard single-echo BUDA-EPI reconstruction pipeline at $R_{\text{in-plane}} \times$ MB = 5 × 2 with 2D pF of 6/8 × 6/8 at 1 × 1 × 5 mm$^3$ slab resolution, where interim SENSE reconstructed blip-up and -down images are fed into FSL TOPUP (49,50) to estimate the field maps. However, due to potential aliasing and g-factor loss at 10-fold total acceleration, the estimated field maps suffer from artifacts, as indicated by the yellow arrows. Fig. 3b shows that PRIME, which has a dual-echo reconstruction pipeline, utilizes the second echo acquired at a lower resolution and in-plane acceleration ($R_{in-plane} \times MB = 4 \times 2$ with 2D pF of $6/8 \times 6/8$) to boost SNR and mitigate artifacts, enabling higher fidelity field map estimation. The two high-resolution shots from the first echo are then reconstructed using S-LORAKS (37,63) with the estimated field maps. The difference images between BUDA-EPI and PRIME (Fig. 3b) show that residual aliasing artifacts in the estimated field maps from BUDA-EPI can hamper distortion correction in the final reconstructed images.

The blip-up and -down shots of the first echo image are jointly reconstructed using a model-based approach with low-rank-based regularization, where the estimated field maps are incorporated in the forward operator. The S-LORAKS is implemented based on the MATLAB (MathWorks, Natick, MA) Toolbox (https://github.com/CongyuLiao/BUDA_SLORAKS) (20,21,37,63). Alternatively, rapid hybrid SENSE reconstruction (51,52) with shot-to-shot phase correction can be used for multi-shot reconstruction using the phase information obtained from the second echo. After reconstructing each gSlider RF-encoded slab, final volume images with the isotropic resolution are reconstructed by resolving RF-encoded slabs with $B_1^+$ and $T_1$ corrections (10,21,68).

To mitigate the motion between the volumes across diffusion directions, especially for mesoscale *in vivo* dMRI at 490 μm isotropic resolution, FSL FLIRT (69,70) is used to register the diffusion volume images using the second echo as navigator. Marchenko-Pastur principal component analysis (MPPCA) algorithm (71–73) was applied using MATLAB implementation (https://github.com/Neurophysics-CFIN/Tensor-MP-PCA/) to denoise the mesoscale data with a patch size of 5. Skull stripping was applied using FSL BET (74) on the b = 0 s/mm$^2$ image. Fractional anisotropy (FA) maps were obtained using FSL DTIFIT (75). Diffusion tractography was analyzed using UKFTractography (https://github.com/pnlbwh/ukftractography) and 3DSlicer (76,77).

**Image Reconstruction Model Comparison**
To validate the DWI reconstruction performance of PRIME, we compared it with SENSE (23) reconstructed blip-up and -down images, hybrid SENSE (51,52), and BUDA-EPI (21). Hybrid SENSE solves a Tikhonov regularized least-squares problem by using the field map and shot-to-shot phase difference estimated from SENSE reconstructions of blip-up and -down shots (51,52), whereas BUDA-EPI used a low-rank modeling of local k-space neighborhoods with phase prior (S-LORAKS) for joint reconstruction of blip-up and -down images (37,63). BUDA-EPI utilized the field maps for its reconstruction, which were estimated from the first echo images at high acceleration, whereas PRIME used the field maps estimated from the second echo images at lower acceleration.

In addition, to evaluate the reconstructed images with a quantitative metric, the 'reference DWIs' were acquired with a total of 10 shots using $R_{in-plane} = 5$ (echo #1) and $R_{in-plane} = 3$ (echo #2) for each blip-up (5 shots) and -down (5 shots) acquisitions. For the second echo, the last 2 shots out of 5 shots for each blip-up and -down acquisition were not used for image reconstruction, since combining 3 shots acquired with $R_{in-plane} = 3$ would already provide an effective $R_{in-plane} = 1$. For PRIME, the first echo images were reconstructed using S-LORAKS with the estimated field maps from the second echo. A root mean square error (RMSE) metric was used for the evaluation of the reconstructed DWIs using $R_{in-plane} = 5$ (echo #1) with 2 shots.

**Diffusion Relaxometry Model**

To demonstrate the PRIME sequence for high-resolution diffusion relaxometry mapping, we utilized a two-compartment tissue model with the spherical mean technique (SMT) (16), while extending the model by accounting for the compartmental $T_2$ values (18,78). This multi-echo SMT (MTE-SMT) model is applicable to study WM microstructure and composition in clinical settings. The MTE-SMT model is included in the package, *Microstructure.jl (79)*, which is developed in the Julia programming language and designed for fast and probabilistic microstructure imaging (https://github.com/Tinggong/Microstructure.jl). Modeling and fitting assumptions for parameter estimation can be flexibly adjusted in Microstructure.jl. In this demonstration, we estimated tissue parameters of the intra-axonal signal fraction ($f_{ia}^0$), intra-axonal $T_2$ ($T_2^{ia}$), extra-cellular $T_2$ ($T_2^{ec}$), parallel diffusivity ($D_∥$), and extra-cellular perpendicular diffusivity ($D_⊥^{ec}$) in the WM, while assuming that parallel diffusivities are the same in intra-axonal and extra-cellular spaces.

**Data Acquisition**

The experiments were conducted with the approval of the Institutional Review Board. Six healthy volunteers were scanned on 3T MAGNETOM Prisma (Siemens Healthineers, Erlangen, Germany) scanner with a 32ch head receive array and Connectome 2.0 (Siemens Healthineers, Erlangen, Germany) scanner with a 72ch head receive array (59,60,80). Imaging sequences were performed on subsets of participants, with the presented figures illustrating data from a single representative subject. The imaging parameters for the PRIME sequence were presented in Supporting Information Table S1, and $B_1^+$ maps were separately acquired using a turbo fast low-angle-shot (turbo-FLASH) sequence (81) with the matched field of view (FOV) at the resolution of 3.44 × 3.44 × 5.5 mm$^3$ to correct the imperfect gSlider RF slab profile due to $B_1^+$ inhomogeneity. δ (small delta) represents diffusion pulse length, and Δ (big delta) represents the time from the beginning of the first gradient to the beginning of the second. For the MTE-SMT diffusion model of dual-echo DWIs, additional b = 0 s/mm$^2$ images were acquired with different TE combinations: [40, 90 ms], [50, 100 ms], [70, 140 ms], and [80, 160 ms].

**RESULTS**

**Image Reconstruction Comparisons**

Fig. 4 shows the reconstructed DWIs for b = 1,000 s/mm$^2$ with $R_{in-plane}$ = 5 and 1D pF (6/8) at 1 × 1 × 4 mm$^3$ resolution. SENSE reconstructed blip-up (1 shot) and -down (1 shot) images show residual aliasing artifacts and amplified noise, and hybrid-SENSE without a phase navigator still suffers from residual artifacts. BUDA-EPI shows better performance than those SENSE-based methods, whereas PRIME outperforms BUDA-EPI and other methods, as presented in the difference images, showing the lowest RMSE values. The estimated field maps using BUDA-EPI and PRIME are presented in Fig. 5. The reference field maps were estimated from the blip-up and -down images acquired using $R_{in-plane}$ = 3 (echo #2) with 6 shots. PRIME enabled higher fidelity of field maps estimation by utilizing the second echo acquired at a lower resolution and in-plane acceleration of $R_{in-plane}$ = 3 to boost SNR and mitigate artifacts. The comparison results using the reference field maps estimated from the first echo are presented in Supporting Information Figure S2.

**Dual-Echo DWI Acquisition**

The reconstructed dual-echo DWIs, for b = 0 and 1,000 s/mm$^2$ with $R_{in-plane}$ × MB = 4 × 2 and 2D pF (6/8 × 6/8) at 1 × 1 × 5 mm$^3$ slab, are presented in Supporting Information Figure S3. The first echo was acquired with $TE_1$ = 53 ms and $ESP_1$ = 0.70 ms, whereas the second echo was acquired with $TE_2$ = 106 ms and $ESP_2$ = 0.70 ms. Both b = 0 and b = 1,000 s/mm² DWIs for both echoes show high geometric fidelity and SNR with high in-plane resolution.

**Multi-Shot Reconstruction with Phase Navigator**

The reconstructed DWIs for b = 1,000 s/mm² with $R_{in-plane}$ = 4 and 2D pF (6/8 × 6/8) at 0.49 × 0.49 × 3.44 mm³ resolution are shown in Fig. 6. Hybrid-SENSE without phase correction suffers from severe aliasing artifacts and shows large phase error compared to S-LORAKS. In contrast, the phase navigator allows Hybrid-SENSE to jointly reconstruct multi-shot DWIs with minimized phase errors, as shown in the second row images. Hybrid-SENSE with a phase navigator demonstrates comparable image qualities to those reconstructed by S-LORAKS while achieving around 100 times faster computation, which is beneficial for reconstructing mesoscale DWIs with a large number of diffusion directions.

**High-Resolution Diffusion Relaxometry**

The high-resolution diffusion relaxometry imaging obtained using the MTE-SMT diffusion model from dual-echo DWIs, which were acquired with 1.1 mm isotropic resolution with $R_{in-plane}$ × MB = 4 × 2 and 2D pF (6/8 × 6/8) for 2-shell acquisitions of b = 0, 750, and 2,000 s/mm² using a clinical scanner, is demonstrated in Supporting Information Figure S4.

In addition, advanced MTE-SMT model results were demonstrated in Fig. 7 using triple-echo DWIs, acquired with 1 mm isotropic resolution with $R_{in-plane}$ × MB = 3 × 2 and 2D pF (6/8 × 6/8) for 3-shell acquisitions of b = 0, 750, 1,200, and 2,000 s/mm² using a Connectome 2.0 scanner. The tissue parameters of the intra-axonal signal fraction ($f_{ia}^0$), intra-axonal $T_2$ ($T_2^{ia}$), extra-cellular $T_2$ ($T_2^{ec}$), parallel diffusivity ($D_\parallel$), and extra-cellular perpendicular diffusivity ($D_\perp^{ec}$) were estimated in the WM. This demonstrated the ability to use PRIME for efficient and high-resolution combined diffusion relaxometry mapping.

**High-Fidelity Mesoscale dMRI**

Fig. 8 shows the mesoscale averaged DWIs and FA maps for b = 1,000 s/mm² with $R_{in-plane}$ × MB = 4 × 2 and 2D pF (6/8 × 6/8) at 490 μm isotropic resolution. By capitalizing on the Connectome 2.0 gradients, $TE_1$ = 49.1 ms was achieved, and the second echo was obtained with $TE_2$ = 130.1 ms using $R_{in-plane}$ × MB = 2 × 2 and 2D pF (6/8 × 6/8) at 2 × 2 × 0.49 mm³ resolution, which were utilized for high-fidelity field maps estimation, phase, and motion navigators. The high-fidelity mesoscale dMRI was achieved only with a single average PRIME data acquisition (number of excitations; NEX = 1) of 90 directions and MPPCA denoising.

A comparison between the acquired and denoised single DWIs for mesoscale data, shown in Fig. 8, is presented in Supporting Information Figure S5. This shows that MPPCA improved the SNR of a single DWI while preserving structure information, as presented in the difference image. Diffusion tractography using PRIME DWIs for b = 1,000 s/mm$^2$ with $R_{in-plane}$ × MB = 4 × 2 and 2D pF (6/8 × 6/8) at 490 μm isotropic resolution is demonstrated in Fig. 9. As shown in the magnified image, detailed u-fiber connectivities are observed with high fidelity in the tractography results.

**DISCUSSION**

In this study, we proposed a new highly accelerated distortion-free dMRI sequence, PRIME, which benefits from incorporating additional echoes without prolonging TR, especially when gSlider RF encoding was used. Through *in vivo* experiments, we demonstrated that PRIME could be utilized for the following cases: 1) high-fidelity DWIs acquisition with a high acceleration factor, 2) high-resolution diffusion relaxometry imaging using MTE-SMT, and 3) mesoscale DWIs of 490 μm isotropic resolution for *in vivo* data.

While BUDA-EPI (21) with gSlider RF encoding has been demonstrated to achieve high-fidelity, high-resolution dMRI, it has limited in-plane acceleration factor per shot and substantially long dead time due to the long TR coming from gSlider RF encoding (10). In particular, residual artifacts in the SENSE reconstructed images were propagated to the estimated field maps when high acceleration per shot was used, e.g., $R_{in-plnae}$ × MB = 5 × 2, whereas PRIME is able to get high-fidelity field maps from the additional low-resolution second echo images without incurring a scan time penalty by using the dead time in each TR. While a single B$_0$ field map acquired from two b = 0 s/mm² volumes at the beginning of a scan may be enough for low b-values, it is insufficient in our context. Our work targets high-resolution (<500 μm resolution), high-acceleration ($R_{in-plane}$ up to 5) dMRI using multi-shot acquisitions, where eddy current-induced distortions vary with diffusion direction and gradient strength. These dynamic distortions cannot be captured by a static B$_0$ field map estimated only at the beginning of the scan. In contrast, PRIME uses the second echo of each shot to estimate dynamic field maps directly from each DWI, enabling accurate correction of both B₀ inhomogeneity and direction-dependent eddy currents. Importantly, Fig. 5 demonstrated that these field maps are highly consistent with reference maps reconstructed from a 10-shot, low-acceleration acquisition.

Multi-shot EPI acquisition reduces voxel blurring and geometric distortion by reducing the readout duration in each shot; however, shot-to-shot phase variations due to physiological noise make it challenging to combine the shots. One of the navigator-free methods, S-LORAKS (37,63), which uses low-rank modeling of local k-space neighborhoods with a smooth phase prior, could reconstruct the high-fidelity DWIs by combining two shots of blip-up and -down images with the incorporated field maps obtained from the second echo. While S-LORAKS could provide high-fidelity images, it required substantial reconstruction time, which took about 20 min per slab at $0.49 \times 0.49 \times 3.44$ mm³ resolution on MATLAB (20). Utilizing the second echo data as a navigator for phase correction between shots made the reconstruction ~100-fold faster by avoiding intensive low-rank-based computations while achieving comparable image qualities, as shown in Fig. 6.

PRIME showed comparable signal levels of the second echo images with those of single-echo images that were separately acquired with another scan, as shown in Supporting Information Figure S1. Using a high time-bandwidth product of the gSlider RF encoding and non-Carr-Purcell-Meiboom-Gill (CPMG) condition with orthogonal crusher gradients could achieve comparable signal levels of the second of PRIME with those of single-echo. This enabled high-resolution triple-echo acquisition of PRIME, which was utilized for diffusion relaxometry mapping with diffusion models.

In PRIME experiments, the effective ESP of the first and second echo readouts was matched to have the same geometric distortion, especially for the shot-to-shot phase correction and field map estimation. If both echoes do not have the matched effective ESP and the $B_0$ correction is not performed, each DWI shot would be similar to the original gSlider sequence; however, it would be difficult to combine two shots acquired with different phase encoding directions without $B_0$ correction and phase navigation. Thus, acquiring multiple echoes with matched effective ESP to estimate shot-to-shot phase difference and field maps enables high-fidelity DWI reconstruction.

While only partial Fourier in the frequency-encoding direction directly reduces ESP, we note that 2D pF, when applied to both frequency and phase-encoding directions, reduces the total EPI readout duration, which allows for an earlier placement of the spin echo and a corresponding reduction in TE. This shorter readout not only helps mitigate $T_2$ and $T_2^*$-related voxel blurring but also improves SNR and reduces the signal dropout, especially in regions with large susceptibility variations. While geometric distortion itself is determined primarily by $B_0$ inhomogeneity and ESP, reduced TE improves the robustness of reconstruction and indirectly benefits image quality and fidelity in high-resolution diffusion imaging. In our implementation, the second echo used for field map estimation is intentionally acquired with lower in-plane acceleration compared to the first echo. This second echo is designed with reduced acceleration, which increases the SNR (due to increased noise averaging via Fourier encoding and reduced g-factor penalty) and improves the conditioning of the interim reconstructions used for field map estimation. Importantly, both echoes are acquired with matched effective ESP to ensure distortion and geometric consistency. This design allows the second echo to provide higher fidelity field maps, even under the constraints of high-resolution and highly accelerated DWI.

Recent advances in MRI hardware, such as Connectome 2.0 (59,60) and MAGNUS (82) scanners, enable the design of more advanced dMRI sequences using strong gradient coils. By capitalizing on the Connectome 2.0 gradients that have a maximum gradient strength of 500 mT/m and a maximum slew rate of 600 T/m/s, we achieved $\delta/\Delta$ = 2.3/20.9 ms for b = 1,000 s/mm$^2$, $TE_1$ = 49.1 ms, $TE_2$ = 130.1 ms, and effective ESP = 0.15 ms for mesoscale 490 μm isotropic resolution. If the same resolution needs to be achieved on a clinical scanner, e.g., a 3T MAGNETOM Prisma scanner which has a maximum gradient strength of 80 mT/m and a maximum slew rate of 200 T/m/s, we need $\delta/\Delta$ = 12.3/46.3 ms for b = 1,000 s/mm$^2$, $TE_1$ = 100 ms, $TE_2$ = 239 ms, and effective ESP = 0.275 ms for mesoscale 490 μm isotropic resolution including consideration of peripheral nerve stimulation limitation. Thus, the Connectome 2.0 scanner allowed us to reduce $TE_1/TE_2$ by about 50.9/45.6 % and effective ESP by about 45.4 %. For a voxel with $T_2$ = 65 ms, for instance, this corresponds to a 118.8 % SNR gain attributable to the reduced TE and 35.4 % SNR loss due to the increased bandwidth, resulting in a net 61.6 % SNR gain. With short TE and ESP, high-fidelity mesoscale DWIs with high SNR and less geometric distortion were obtained using the PRIME sequence and reconstruction scheme. Here, we used $R_{in-plane}$ of 4-fold instead of 5-fold to retain SNR for mesoscale diffusion imaging. Increasing the in-plane acceleration can reduce TE by reducing the number of frequency-encoding lines while it reduces Fourier encoding as well; thus, there is a trade-off between TE reduction and decreased noise averaging due to Fourier encoding.

PRIME achieved high-fidelity mesoscale DWI at 490 μm isotropic resolution. If gSlider RF encoding is not employed in the framework, the achievable spatial resolution of PRIME can be 0.49 × 0.49 × 1.30 mm³ while achieving similar SNR to the ones presented in Fig. 9. Since employing gSlider RF encoding gives an SNR boost of ~√(# of gSlider RFs), achievable voxel size (in particular for slice direction) with the same SNR can be 0.49 × √7 = 1.30 mm. This spatial resolution can be identical to 0.678 mm isotropic resolution.

Though diffusion models can extract microstructure features of the human brain from dMRI, the resolutions of *in vivo* data are mostly limited to 1.5~2.5 mm isotropic resolution (13–18). Here, we demonstrated that the PRIME sequence enabled high-resolution (e.g., 1 mm isotropic resolution with a Connectome 2.0 scanner) diffusion relaxometry using the WM diffusion model, MTE-SMT, which provided diffusion relaxometry parameters, including intra-axonal signal fraction ($f_{ia}^0$), intra-axonal $T_2$ ($T_2^{ia}$), extra-cellular $T_2$ ($T_2^{ec}$), parallel diffusivity ($D_\parallel$), and extra-cellular perpendicular diffusivity ($D_\perp^{ec}$). Triple-echo DWIs were obtained with TEs = 40, 84, and 128 ms with a single sequence, demonstrating the effectiveness of the multi-echo PRIME acquisition.

PRIME will lend itself to more advanced diffusion models, such as neurite orientation dispersion and density imaging (NODDI) (13), soma and neurite density imaging (SANDI) (14), neurite exchange imaging (NEXI) (15), and their multi-TE extensions (18,83,84). In particular, PRIME may enable advanced gray matter (GM) analysis, where current GM models use around 2 mm isotropic resolutions, which are too coarse to image the cortex (1~4 mm thickness) due to severe partial volume effects. By using the VERSE technique, triple-echo PRIME was achieved without SAR constraints, where a VERSE factor of 2.2 was used for MB = 2. PRIME took advantage of long TR, which was needed for gSlider RF encoding to mitigate spin history-related slab boundary artifacts (10). Though $B_1^+$ and $T_1$ corrections could reduce the slab boundary artifacts (68), PRIME showed residual artifacts with a TR of around 3.5~4.6 s. Increasing TR may mitigate the slab boundary artifacts while reducing the scan efficiency and allowing more dead time. According to the previous studies, TR of 3.5 s provided a good balance in the trade-off between SNR efficiency versus spin history and motion sensitivity (21,68). The current MB factor of 2 already allows us to acquire triple echoes within TR of 3.5 s, especially for 1 mm isotropic resolution of triple-echo DWIs for diffusion relaxometry with MTE-SMT model utilizing Connectome 2.0 scanner. Employing MB > 2 may trigger RF voltage clipping and SAR constraints, requiring longer TR in that case. Reducing the number of diffusion directions using gSlider-spherical ridgelets (gSlider-SR) (9) can be one way to reduce the total scan time while maintaining SNR and angular information. Super-resolution techniques with rotating views (56,85,86) can be alternative strategies to achieve high-resolution dMRI while minimizing slab boundary artifacts, yet these necessitate motion correction across the rotating views since these are acquired several minutes apart.

An open-source and vendor-neutral MR sequence development framework, *Pulseq (64)*, was used to design and implement the sequence, where various sequences, including spiral, CEST, quantitative MRI, and dMRI (87–91), have been designed using it. By capitalizing on the open-source framework, PRIME will also lend itself to harmonizing the sequences across different scanners, sites, or vendors, such as the high-performance MAGNUS system.

## CONCLUSION

We proposed a pulse sequence of phase-reversed interleaved multi-echo acquisition called PRIME for highly accelerated, high-resolution, and distortion-free dMRI, which enabled mesoscale dMRI using high-fidelity field maps as well as high-resolution diffusion relaxometry imaging, by inserting additional echoes without prolonging scan time.

## ACKNOWLEDGMENTS

This work was supported by research grants NIH R01 EB032378, R01 EB028797, R01MH132610, R01MH125860, R01MH116173, R01HD114719, R01EY030434, P41 EB030006, U01 EB026996, UG3 EB034875, R21 AG082377, R21 EB036105, NVIDIA Corporation for computing support, and National Research Foundation of Korea (NRF) grant funded by the Korea government (MSIT) (RS-2025-00555277). MES received funding from Max Planck School of Cognition, supported by the German Federal Ministry of Education and Research (BMBF) and the Max Planck Society.

## DATA AVAILABILITY STATEMENT

The source codes can be found here: https://github.com/yohan-jun/PRIME

**Figure Legends**

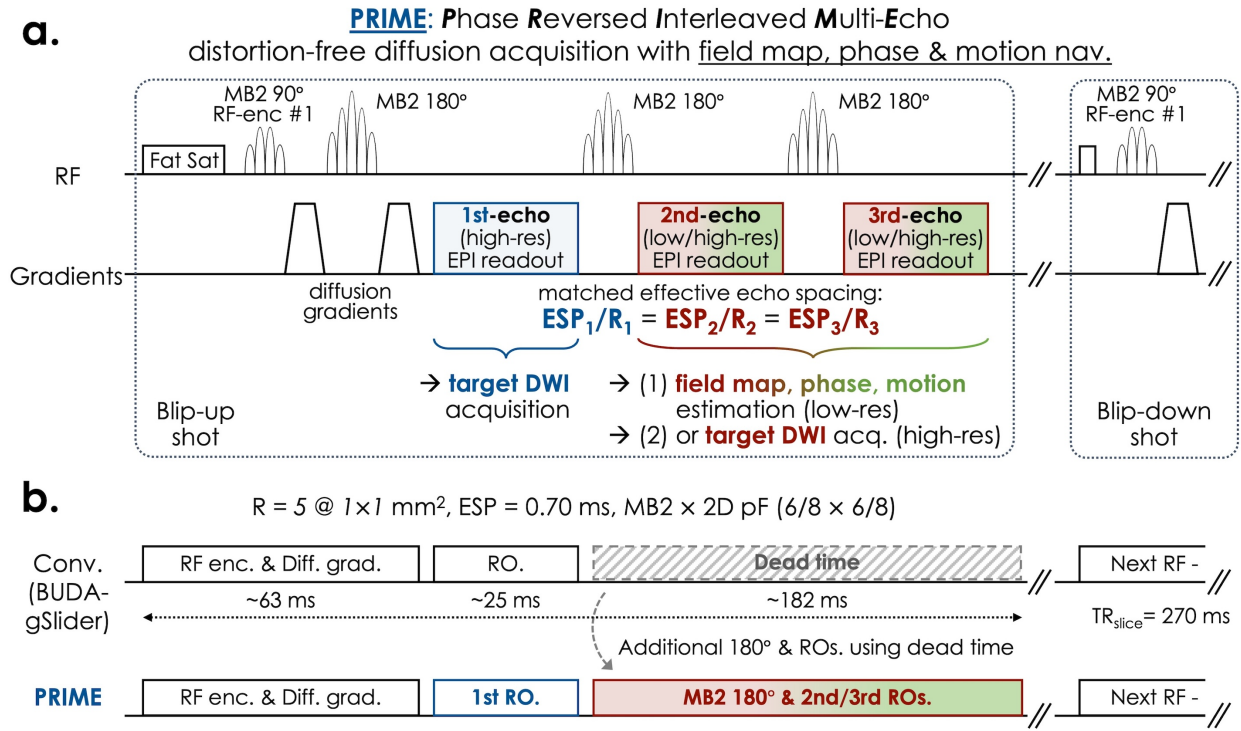

**Figure 1.** (a) Pulse sequence diagram of Phase Reversed Interleaved Multi-Echo acquisition (PRIME) for highly accelerated distortion-free diffusion MRI, which has two or three spin echoes where the first echo is for target diffusion-weighted imaging (DWI) acquisition with high-resolution and the second or third echo is acquired with either 1) lower-resolution for field map estimation along with phase and motion navigator, or 2) matching-resolution for the second or third echo of target DWI acquisition. The effective echo spacing (ESP) of the first, second, and third echo readouts is matched to have the same geometric distortion. (b) Sequence timing comparison between conventional blip-up and -down (BUDA)-EPI and PRIME. Single-echo acquisition of BUDA-EPI with generalized slice dithered enhanced resolution (gSlider) radiofrequency (RF) encoding has a long dead time in each TR to mitigate spin history-related slab boundary artifacts, whereas PRIME utilizes this time by inserting additional refocusing pulse and readouts without prolonging TR.

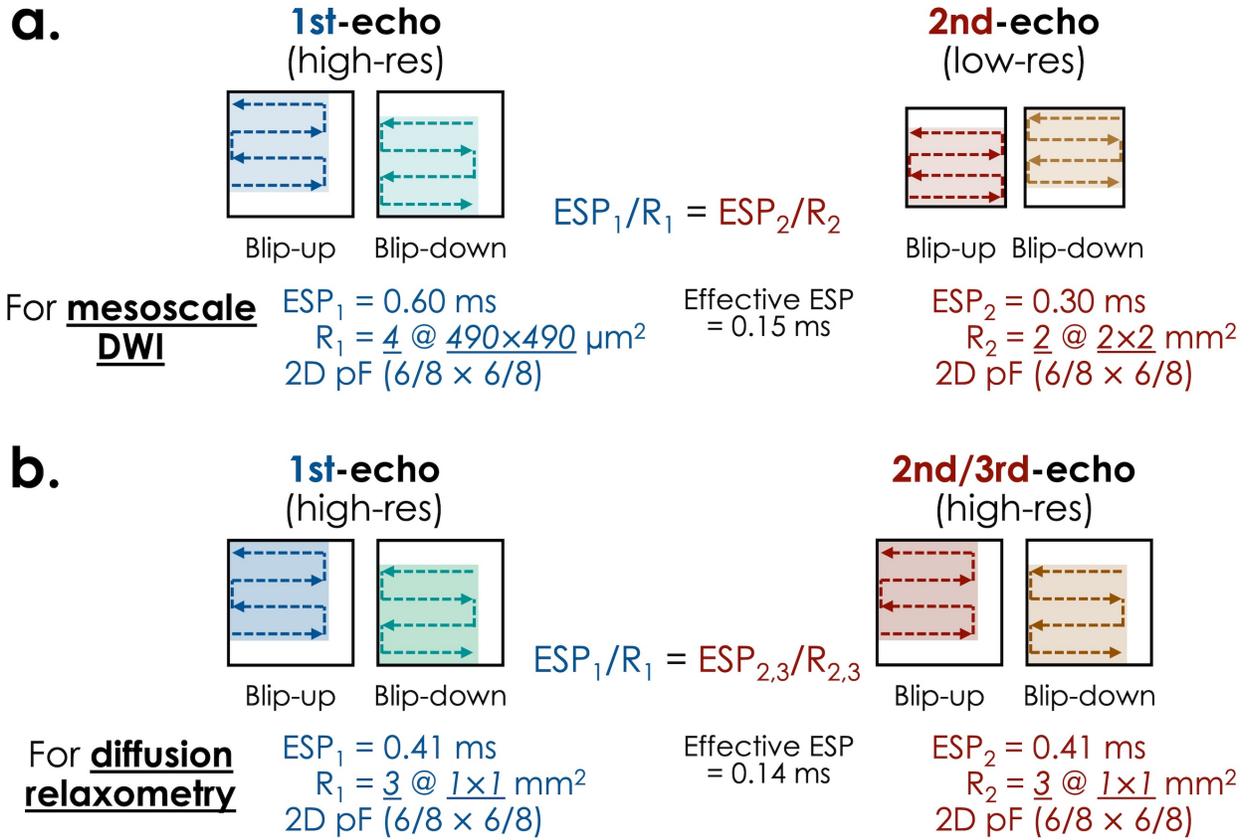

**Figure 2.** Schematic k-space trajectories of PRIME. (a) For high-fidelity field maps estimation, the first echo is acquired with high in-plane acceleration, whereas the second echo is acquired with the same effective ESP as the first echo for lower in-plane acceleration and resolution. Mesoscale diffusion imaging can be achieved with the high gradient strength of the Connectome 2.0 scanner. (b) For high-resolution diffusion relaxometry, all echoes have the same resolution and in-plane acceleration to match the voxel size and distortion.

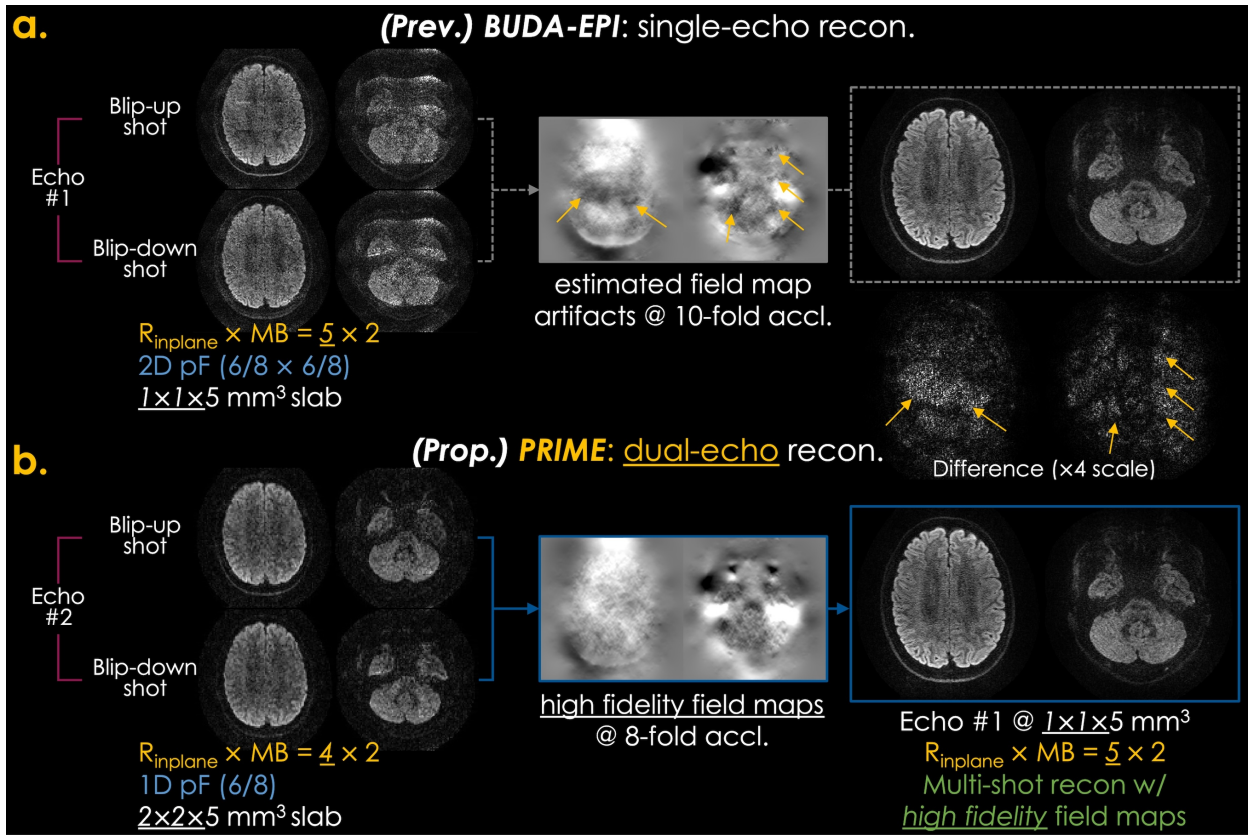

**Figure 3.** (a) Standard single-echo blip-up and -down (BUDA)-EPI reconstruction pipeline, where interim SENSE reconstructed blip-up and -down images are fed into FSL TOPUP to estimate the field maps. However, due to aliasing artifacts and g-factor loss at 10-fold acceleration, the estimated fieldmaps suffer from artifacts, as indicated by the yellow arrows. (b) PRIME has a dual-echo reconstruction pipeline, which utilizes the second echo acquired at a lower resolution and in-plane acceleration to boost SNR and mitigate artifacts, enabling higher fidelity of field maps estimation. The two high-resolution shots from the first echo are then jointly reconstructed using the estimated field maps. Note that single-direction DWI for b = 1,000 s/mm$^2$ acquired within 35 sec per direction is shown here.

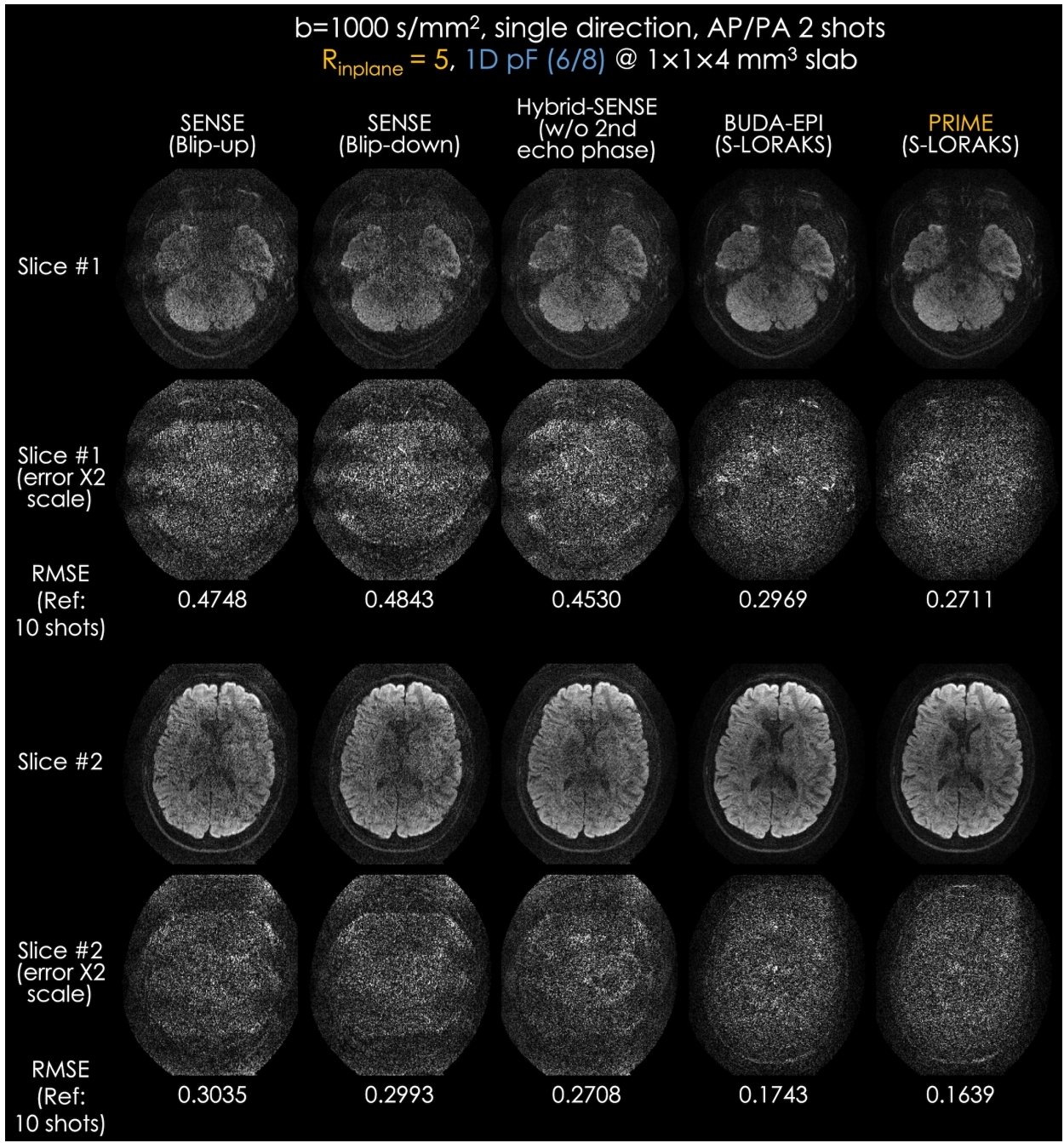

**Figure 4.** Reconstructed DWIs using 2 shots of blip-up and -down acquisitions for b = 1,000 s/mm² with $R_{in-plane}$ = 5 and 1D pF (6/8) at 1 × 1 × 4 mm³ resolution using SENSE, Hybrid-SENSE, blip-up and -down (BUDA)-EPI, and PRIME. The reference DWIs were acquired with a total of 10 shots using $R_{in-plane}$ = 5 (echo #1) and 6 shots using $R_{in-plane}$ = 3 (echo #2) with blip-up and -down acquisitions, and they were reconstructed using S-LORAKS with the estimated field maps from the second echo. A root mean square error (RMSE) metric was used for the evaluation of the reconstructed DWIs using $R_{in-plane}$ = 5 (echo #1) with 2 shots. Note that single-direction DWI for b = 1,000 s/mm$^2$ acquired within 9.7 sec (for 2-shots) and 48.5 sec (for 10-shots reference) per direction is shown here.

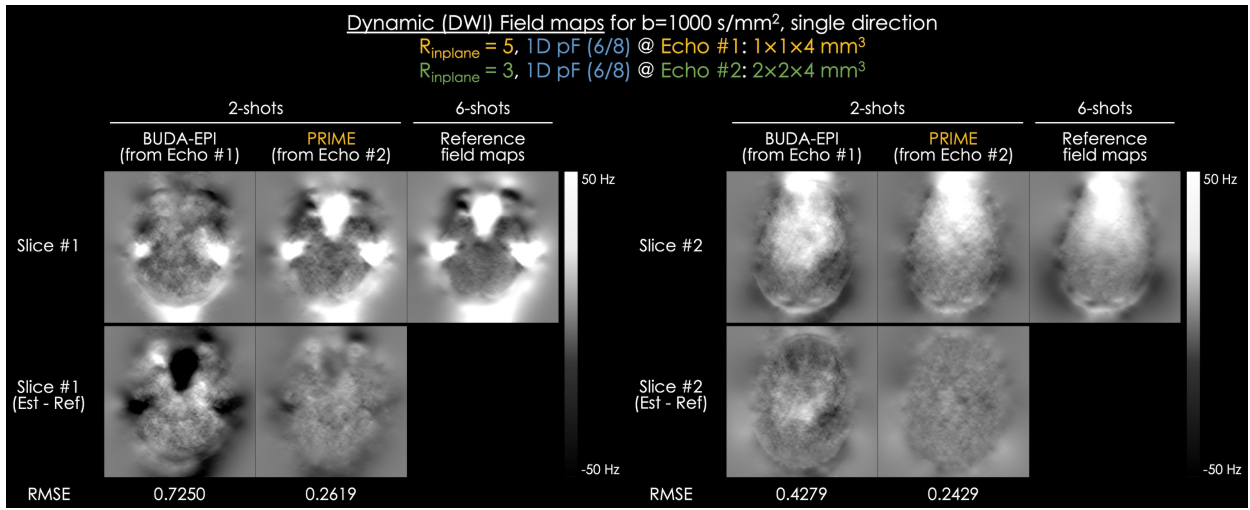

**Figure 5.** Estimated field maps for b = 1,000 s/mm$^2$ with $R_{\text{in-plane}}$ = 5 and 1D pF (6/8) at $1 \times 1 \times 4$ mm$^3$ resolution (echo #1) and $R_{\text{in-plane}}$ = 3 and 1D pF (6/8) at $2 \times 2 \times 4$ mm$^3$ resolution (echo #2) using blip-up and -down (BUDA)-EPI and PRIME. The reference field maps were estimated from the blip-up and -down images acquired using $R_{\text{in-plane}}$ = 3 (echo #2) with 6 shots. A root mean square error (RMSE) metric was used for the evaluation of the estimated field maps compared to the references. Note that field maps of single-direction DWI for b = 1,000 s/mm$^2$ acquired within 9.7 sec (for 2-shots) and 29.1 sec (for 6-shots reference) per direction are shown here.

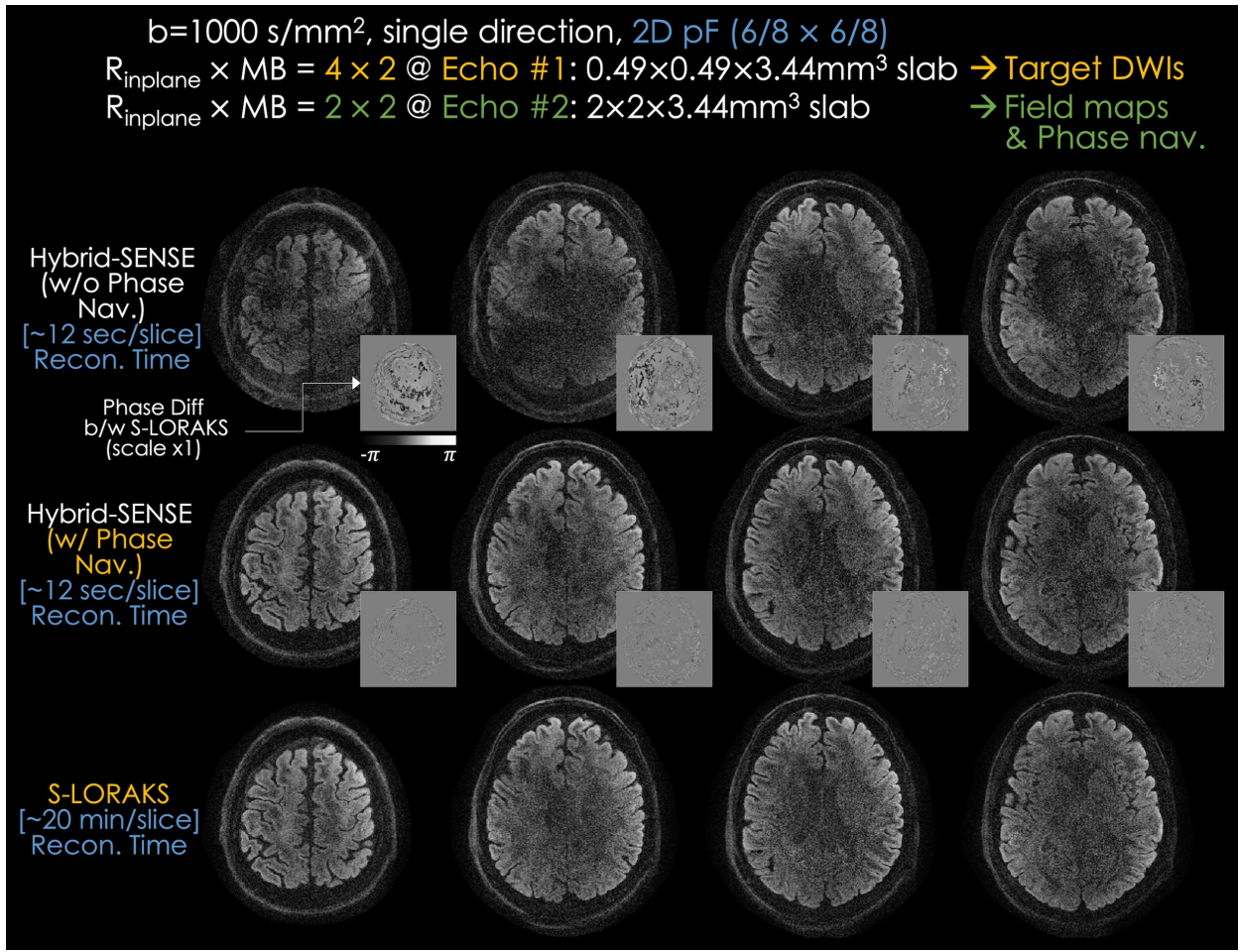

**Figure 6.** Reconstructed DWIs acquired with PRIME for b = 1,000 s/mm$^2$ with $R_{in-plane}$ = 4 and 2D pF (6/8 × 6/8) at 0.49 × 0.49 × 3.44 mm$^3$ resolution using Hybrid-SENSE with and without phase navigator and S-LORAKS. Compared to Hybrid-SENSE without a phase navigator, Hybrid-SENSE with a phase navigator shows comparable magnitude images and reduced phase error with S-LORAKS, while achieving around 100 times faster computation time. Note that single-direction DWI for b = 1,000 s/mm$^2$ acquired within 9.2 sec per direction and per slab is shown here.

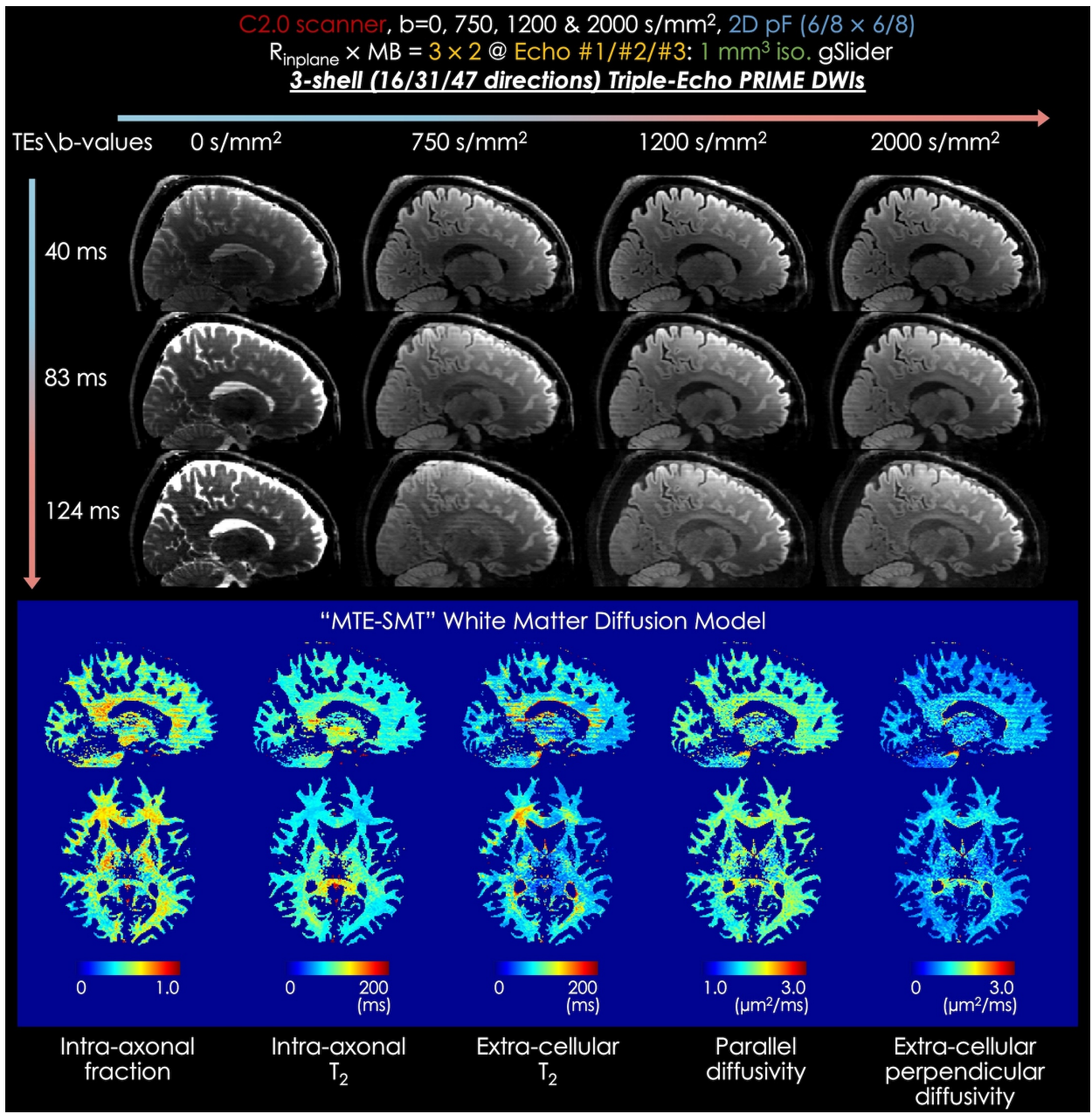

**Figure 7.** High-resolution diffusion relaxometry imaging was obtained using multi-TE spherical mean technique (MTE-SMT) diffusion model from triple-echo PRIME data, which were acquired with 1 mm isotropic resolution with $R_{in-plane} \times MB = 3 \times 2$ and 2D pF (6/8 × 6/8) for 3-shell acquisitions (b = 0, 750, 1200, and 2,000 s/mm$^2$). MTE-SMT provided diffusion relaxometry parameters, including intra-axonal fraction, intra-axonal $T_2$, extra-cellular $T_2$, parallel diffusivity, and extra-cellular perpendicular diffusivity maps. Note that the total scan time for 3-shell acquisitions was 59 min.

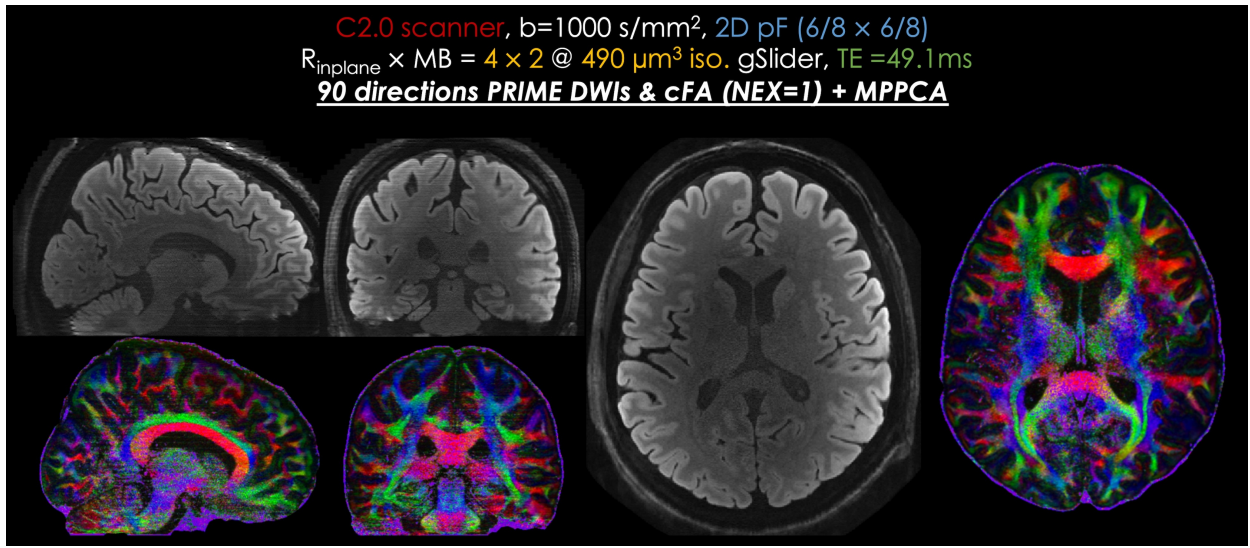

**Figure 8.** Mesoscale averaged diffusion-weighted images and fractional anisotropy maps for b = 1,000 s/mm$^2$ with $R_{\text{in-plane}}$ × MB = 4 × 2 and 2D pF (6/8 × 6/8) at 490 μm isotropic resolution. High-fidelity mesoscale images could be obtained only with a single PRIME data acquisition of 90 directions using a Connectome 2.0 scanner and Marchenko-Pastur principal component analysis (MPPCA) algorithm. Note that the total scan time for single-shell acquisition was 104 min.

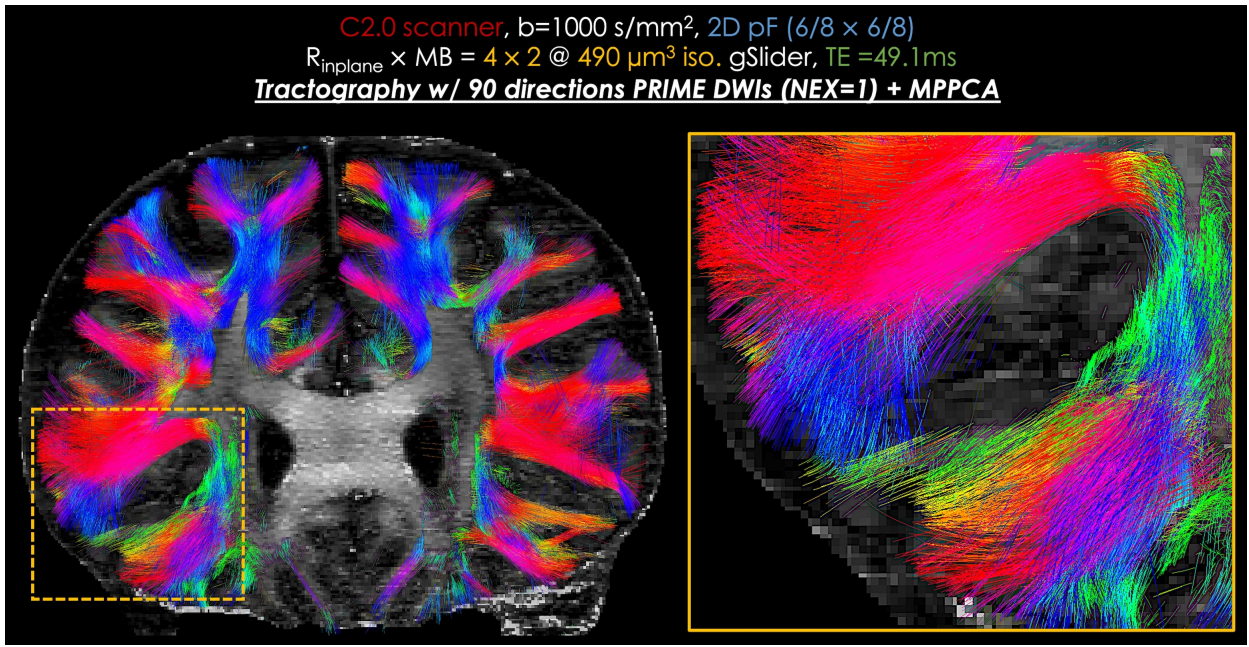

**Figure 9.** Diffusion tractography using PRIME DWIs for b = 1,000 s/mm² with $R_{in\text{-}plane} \times MB = 4 \times 2$ and 2D pF (6/8 × 6/8) at 490 μm isotropic resolution. High-fidelity tractography could be obtained only with a single PRIME data acquisition of 90 directions using a Connectome 2.0 scanner and Marchenko-Pastur principal component analysis (MPPCA) algorithm. Note that the total scan time for single-shell acquisition was 104 min.

**Supporting Information Data**

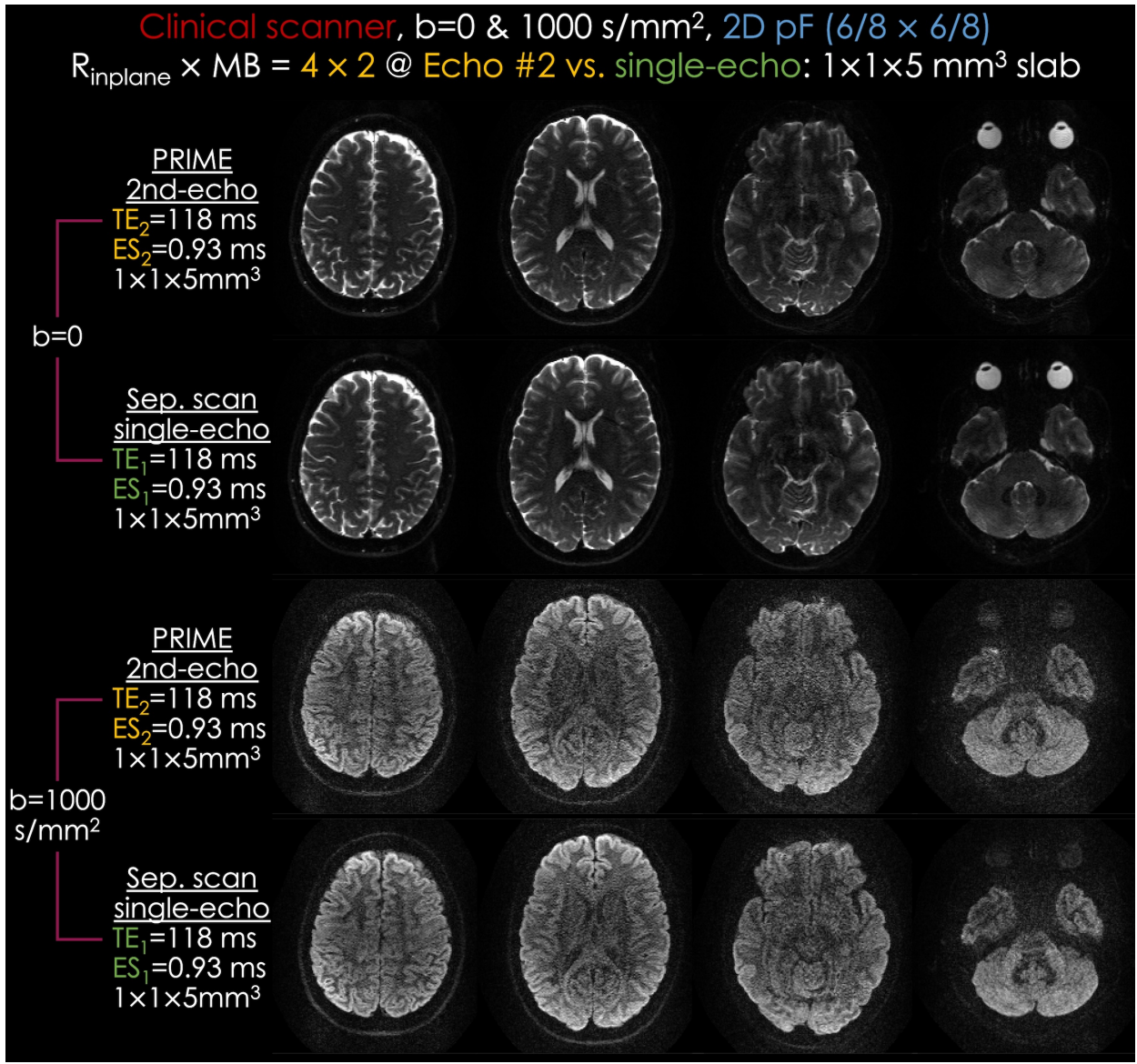

**Supporting Information Figure S1.** Reconstructed second echo images of PRIME and single-echo image separately acquired with the same TE, for b = 0 and 1,000 s/mm² with $R_{in\text{-}plane} \times MB$ = 4 × 2 and 2D pF (6/8 × 6/8) at 1 × 1 × 5 mm³ slab. The single-echo images were acquired separately from another scan with the matched TE = 118 ms. Note that single-direction DWI for b = 0 and 1,000 s/mm² acquired within 7.0 sec per direction and per slab is shown here.

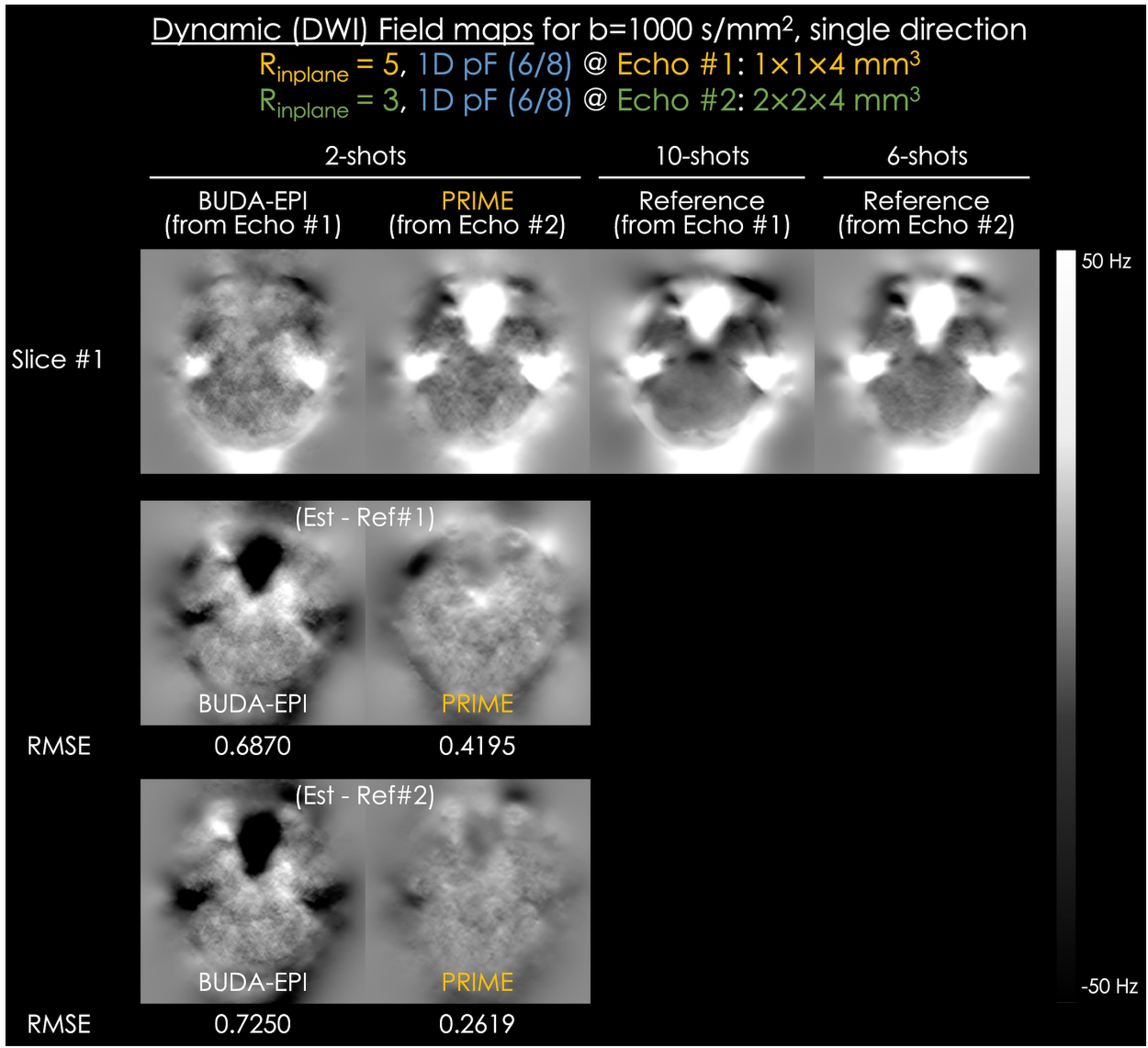

**Supporting Information Figure S2.** Estimated field maps for b = 1,000 s/mm² with $R_{\text{in-plane}} = 5$ and 1D pF (6/8) at $1 \times 1 \times 4$ mm³ resolution (echo #1) and $R_{\text{in-plane}} = 3$ and 1D pF (6/8) at $2 \times 2 \times 4$ mm³ resolution (echo #2) using blip-up and -down (BUDA)-EPI and PRIME. The reference field maps were estimated from the blip-up and -down images acquired using $R_{\text{in-plane}} = 5$ (echo #1) with 10 shots and $R_{\text{in-plane}} = 3$ (echo #2) with 6 shots. A root mean square error (RMSE) metric was used for the evaluation of the estimated field maps compared to the references. Note that field maps of single-direction DWI for b = 1,000 s/mm2 acquired within 9.7 sec (for 2-shots), 29.1 sec (for 6-shots reference), and 48.5 sec (for 10-shots reference) per direction are shown here.

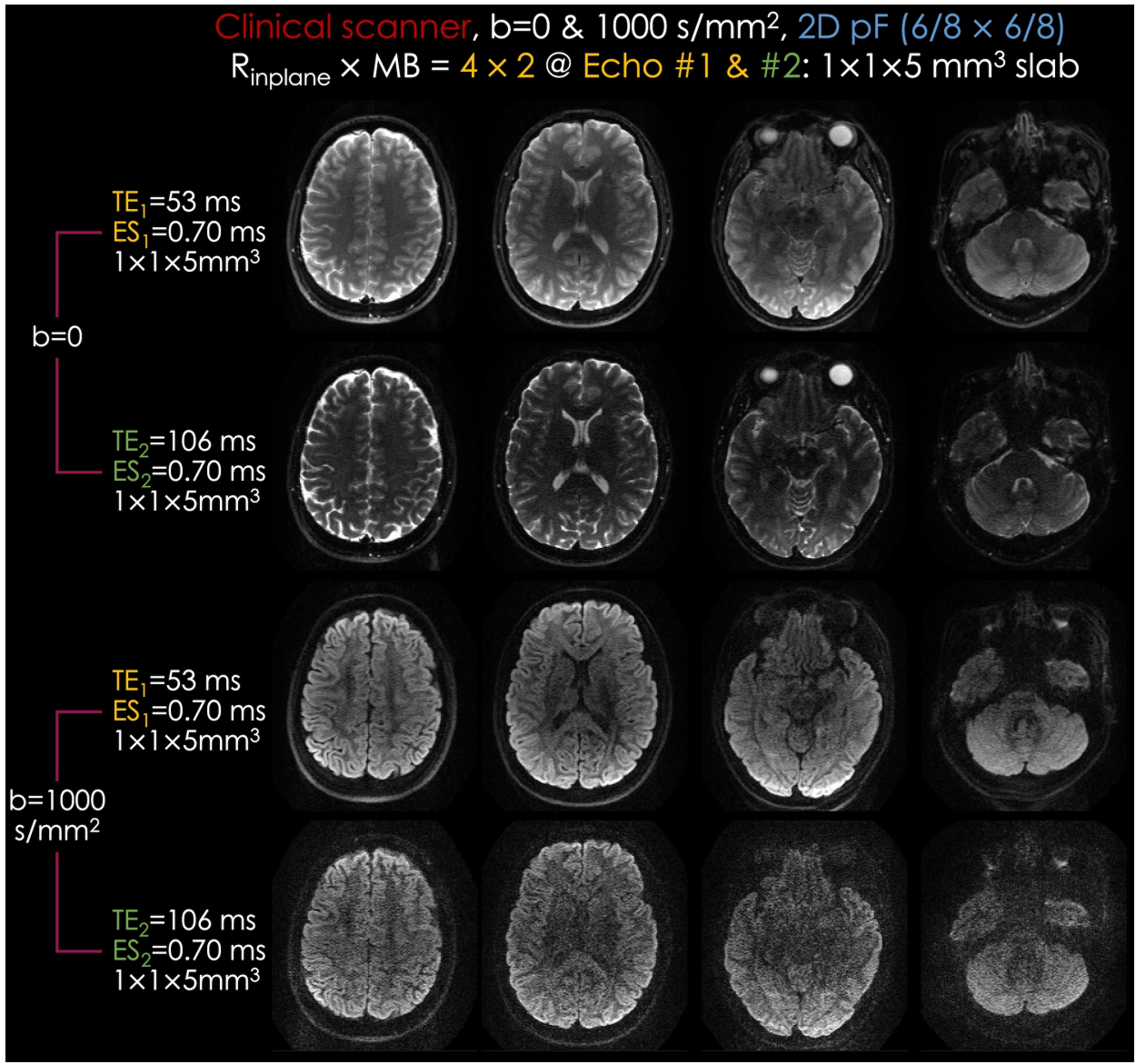

**Supporting Information Figure S3.** Reconstructed dual-echo diffusion-weighted images, for b = 0 and 1,000 s/mm² with $R_{\text{in-plane}} \times MB = 4 \times 2$ and 2D pF (6/8 × 6/8) at 1 × 1 × 5 mm³ slab. The first echo was acquired with $TE_1 = 53$ ms and $ESP_1 = 0.70$ ms, whereas the second echo was acquired with $TE_2 = 106$ ms and $ESP_2 = 0.70$ ms. Both b = 0 and b = 1,000 s/mm² images for both echoes show high geometric fidelity and SNR with high in-plane resolution. Note that single-direction DWI for b = 0 and 1,000 s/mm² acquired within 7.0 sec per direction and per slab is shown here.

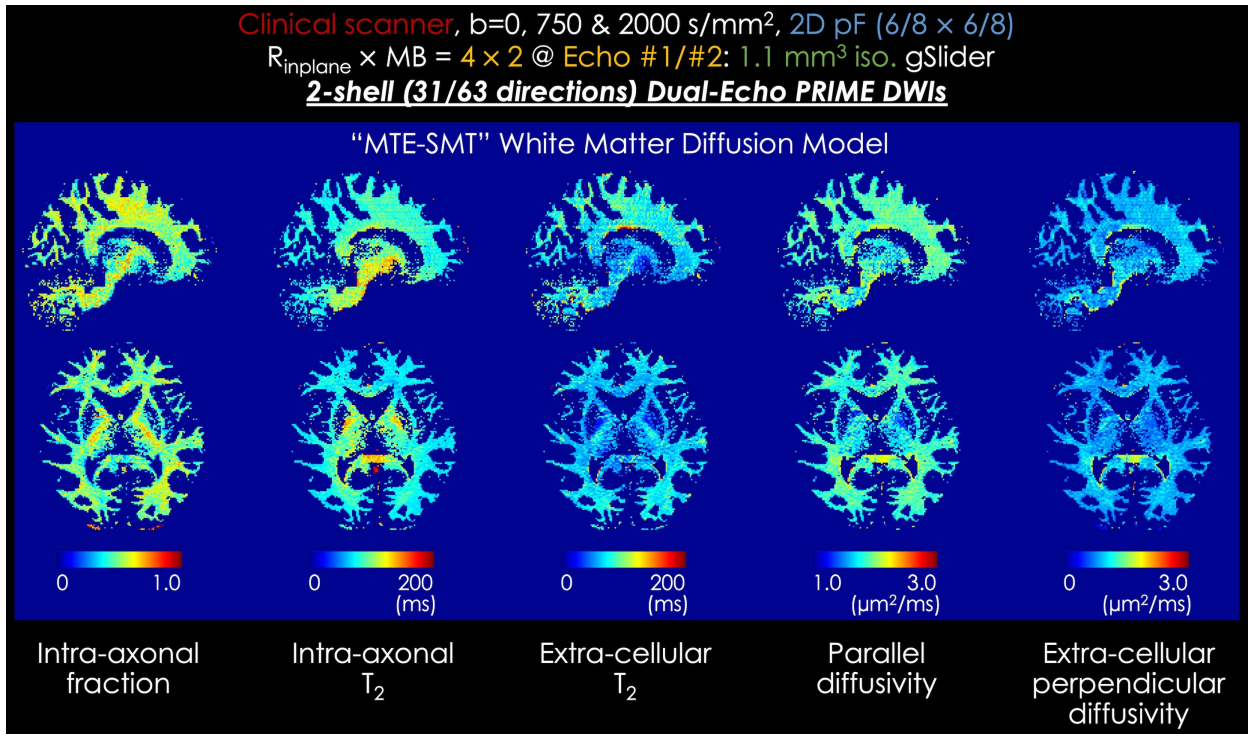

**Supporting Information Figure S4.** High-resolution diffusion relaxometry imaging was obtained using multi-TE spherical mean technique (MTE-SMT) diffusion model from dual-echo PRIME data, which were acquired with 1.1 mm isotropic resolution with $R_{\text{in-plane}} \times MB = 3 \times 2$ and 2D pF (6/8 × 6/8) for 2-shell acquisitions (b = 0, 750, and 2,000 s/mm$^2$). MTE-SMT provided diffusion relaxometry parameters, including intra-axonal fraction, intra-axonal T$_2$, extra-cellular T$_2$, parallel diffusivity, and extra-cellular perpendicular diffusivity maps. Note that the total scan time for 2-shell acquisitions was 61 min.

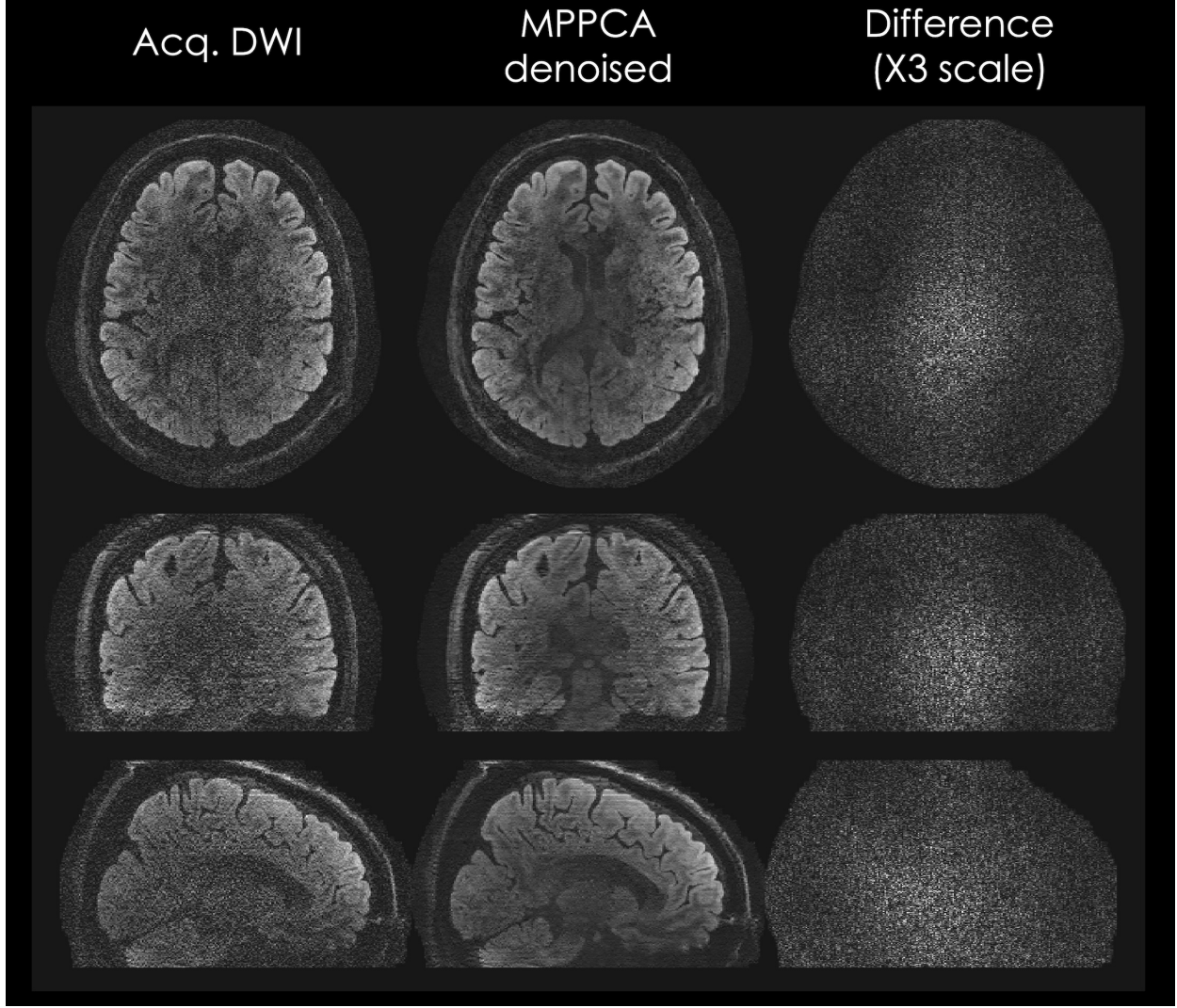

**Supporting Information Figure S5.** Comparison between the acquired and denoised single direction diffusion-weighted images for b = 1,000 s/mm² with $R_{\text{in-plane}} \times MB = 4 \times 2$ and 2D pF (6/8 × 6/8) at 490 μm isotropic resolution. The Marchenko-Pastur principal component analysis (MPPCA) algorithm was used to denoise the images. Note that single-direction DWI for b = 1,000 s/mm² acquired within 46 sec per direction is shown here.

**Supporting Information Table S1.** MRI scan parameters of PRIME sequence used for *in vivo* experiments.

| | High-fidelity field maps acquisition | Multi-shot acquisition for reference image | High-resolution dual-echo DWIs | Diffusion relaxometry with dual-echo DWIs | Diffusion relaxometry with triple-echo DWIs | Mesoscale DWIs |
|---|---|---|---|---|---|---|
| Related Figures | Fig. 3 | Figs. 4, 5, S2 | Figs. S1, S3 | Fig. S4 | Fig. 7 | Figs. 6, 8, 9, S5 |
| FOV | 220×220×130 mm³ | 220×220×130 mm³ | 220×220×130 mm³ | 220×220×143 mm³ | 220×220×130 mm³ | 220×220×130.6 mm³ |
| # of echoes | 2 | 2 | 2 | 2 | 3 | 2 |
| Resolution (Echo #1 / #2 / #3) | 1 mm isotropic 2×2×1 mm³ | 1×1×4 mm³ 2×2×4 mm³ | 1 mm isotropic | 1.1 mm isotropic | 1 mm isotropic | 490 µm isotropic 2×2×0.49 mm³ |
| $R_{in\text{-}plane}$ (Echo #1 / #2 / #3) | 5 / 4 | 5 / 3 | 4 | 4 | 3 | 4 / 2 |
| # of shots (Echo #1 / #2 / #3) | 2 / 2 | 10 / 6 | 2 | 2 | 2 | 2 |
| pF (Echo #1 / #2 / #3) | 6/8 × 6/8 6/8 | 6/8 | 6/8 × 6/8 | 6/8 × 6/8 | 6/8 × 6/8 | 6/8 × 6/8 |
| MB | 2 | 1 | 2 | 2 | 2 | 2 |
| TR | 3,500 ms | 4,850 ms | 3,500 ms | 3,500 ms | 3,500 ms | 4,600 ms |
| $TE_1$ / $TE_2$ / $TE_3$ | 54 / 108 ms | 51 / 102 ms | 53 / 106 ms | 60 / 120 ms | 40 / 84 / 126 ms | 49.1 / 130.1 ms |
| $ESP_1$ / $ESP_2$ / $ESP_3$ | 0.70 / 0.56 ms | 0.95 / 0.57 ms | 0.70 ms | 0.67 ms | 0.41 ms | 0.60 / 0.30 ms |
| δ / Δ | 14.0 / 23.7 ms | 11.3 / 23.9 ms | 11.1 / 23.2 ms | 16.2 / 26.8 ms | 7.4 / 16.8 ms | 2.3 / 20.9 ms |
| b-values | b = 0 (1 rep), 1,000 s/mm² (6 dirs.) | b = 0 (4 reps.), 1,000 s/mm² (32 dirs.) | b = 0 (1 rep), 1,000 s/mm² (6 dirs.) | b = 0 (6 reps.), b = 0 (4 different TEs), 750 s/mm² (31 dirs.), 2,000 s/mm² (63 dirs.) | b = 0 (7 reps.), 750 s/mm² (16 dirs.), 1,200 s/mm² (31 dirs.), 2,000 s/mm² (47 dirs.) | b = 0 (7 reps.), 1,000 s/mm² (90 dirs.) |
| Scanner | 3T Prisma | 3T Prisma | 3T Prisma | 3T Prisma | 3T Connectome 2.0 | 3T Connectome 2.0 |
| Total Scan Time | 3.5 min | 30 min | 3.5 min | 61 min | 59 min | 104 min |